\documentclass[12pt]{iopart}

\usepackage{iopams}
\usepackage{graphicx}
\usepackage{stmaryrd}
\expandafter\let\csname equation*\endcsname\relax
\expandafter\let\csname endequation*\endcsname\relax
\usepackage{mathtools}
\usepackage[dvipsnames]{xcolor}
\renewcommand{\vec}[1]{\boldsymbol{#1}}

\newcommand{\caB}{{\mathcal B}}

\newcommand{\caH}{{\mathcal H}}

\newcommand{\caO}{{\mathcal O}}

\newcommand{\caS}{{\mathcal S}}
\newcommand{\caT}{{\mathcal T}}

\newcommand{\caV}{{\mathcal V}}

\newcommand{\diff}{d} 

\newcommand{\mean}[1]{{\left< #1 \right>}}
\newcommand{\genL}{{\mathbb L}}
\newcommand{\genW}{{\mathbb W}}

\begin{document}

\title{Mesoscopic virial equation for nonequilibrium statistical mechanics}

\author{G Falasco$^{1}$, F Baldovin$^{2,3,4}$, K Kroy$^5$, and M Baiesi$^{2,3}$}

\address{$^1$Max Planck Institute for Mathematics in the Sciences, Inselstr. 22,
04103 Leipzig, Germany}

\address{$^2$ Dipartimento di Fisica ed Astronomia, Universit\`a di Padova, Via Marzolo 8, I-35131 Padova, Italy}

\address{$^3$ INFN, Sezione di Padova, Via Marzolo 8, I-35131 Padova, Italy}

\address{$^4$ CNISM - Sezione di Padova,  Via Marzolo 8, I-35131 Padova, Italy}

\address{$^5$ Institut f\"ur Theoretische Physik, Universit\"at Leipzig,  Postfach 100 920, D-04009 Leipzig, Germany}

\ead{gianmaria.falasco@mis.mpg.de}
\ead{baldovin@pd.infn.it}
\ead{klaus.kroy@itp.uni-leipzig.de}
\ead{baiesi@pd.infn.it}


\begin{abstract}We derive {a class of} mesoscopic virial equations governing energy partition between conjugate position and momentum variables of individual degrees of freedom.  {They are shown to apply to a wide range} of nonequilibrium steady states with stochastic (Langevin) and deterministic (Nos\'e--Hoover) dynamics, and to extend to collective modes for models of heat-conducting lattices. 
{A generalised macroscopic virial theorem ensues upon summation over all degrees of freedom. 
This theorem allows for the derivation of nonequilibrium state equations
that involve} dissipative heat flows on the same footing with state variables, as exemplified for inertial Brownian motion with solid friction and {overdamped} active Brownian particles subject to inhomogeneous pressure.
\end{abstract}

\maketitle

\section{Introduction}
From equilibrium statistical mechanics we are
accustomed to the idea that
there is energy equipartition among all quadratic degrees of freedom
of classical systems, and that the ``energy bit''
corresponds to $k_{\rm B} T/2$, half of the temperature times the
Boltzmann constant.  While momenta usually appear with the quadratic
contribution of the kinetic energy in the Hamiltonian $\caH$, for a
position variable $q_i$ one has more generally that it is the average of 
$q_i \partial_{ q_i} \caH$ which equals the energy bit.
The sum over all degrees of freedom yields the
virial theorem~\cite{gallavotti,mar07}, which connects the average total
kinetic energy with the term $\sum_i \langle q_i \partial_ { q_i} \caH \rangle$
named virial by Clausius.

Out of equilibrium, the equipartition of energy is not granted.
Indeed, recent
experiments with heat-conducting metals show intriguing deviations from
equipartition, related to enhancements of low-frequency vibrational
modes that may become even ``hotter'' than the highest boundary
temperature~\cite{con13}. Similar deviations from
equipartition are observed for strongly heated cantilevers~\cite{agu15}
and Brownian particles~\cite{jol11, fal14}.
These are some out of many manifestations of
nontrivial effects characterizing systems driven far from thermodynamic
equilibrium. 
They imply the need for a critical revision of
statistical mechanics equilibrium results,  with the aim of pointing out
possible extensions to nonequilibrium conditions. 

In this work we discuss a generalization of the equipartition theorem,
formulated in the context of modern nonequilibrium physics. It
takes the form of 
{\em mesoscopic virial equations} (MVEs), involving kinetic and
dynamical aspects specific to pairs of momentum-position conjugate
variables. A MVE determines how thermal energy is distributed 
between any such pair of variables. 
For Langevin dynamics, we discuss both the inertial and the
overdamped versions of the equation; 
the former is easily extended to cover
Nos\'e-Hoover dynamics for thermostated simulations.  
Summation of a MVE over all degrees of freedom generates the virial
theorem, which we discuss also for the case of explicitly
nonconservative forces. 
That the virial theorem holds beyond thermal equilibrium should not come as a surprise, since it is a result derivable in
classical mechanics without appealing to statistical arguments~\footnote{In classical mechanics the theorem involves time averages, which are customarily exchanged with probability averages under the ergodic assumption \cite{fal15}.}. 
The simple mathematical derivations we employ are slightly different from
the conventional line of arguments dating back to Chandrasekhar's
work~\cite{cha64,mar07}. The main novelty of our approach is that we
work consistently in the context of nonequilibrium systems, 
and that our derivations easily carry over to
deterministic thermostats. Moreover, we characterise 
energy partition even for collective macroscopic variables,
such as single normal modes, out of equilibrium.
We further show that our results allow for the derivation of 
equations of state for nonequilibrium steady states. 
As an illustrative example, we provide a full derivation of the pressure equation for a 
{well-known} model of active matter~\cite{sol15}.

\section{Langevin dynamics}\label{sec:Langevin}
Consider $N$ interacting particles evolving in $d$ dimensions, with
generalised coordinates $\{ q_i, p_i\}$, $i=1,\dots,N d$. Each degree
of freedom has mass $m_i$ and the total energy is given by the
Hamiltonian
\begin{equation}
\label{H}
\caH= \sum_{i=1}^{N d} \frac{ p_i^2}{2m_i} + U(\{ q_i\})\, ,
\end{equation}
where $U( \{ q_i\})$ contains a confining potential energy that allows the
system to reach a stationary state in the absence of external,
time-dependent driving. In addition, nonconservative forces $f_i$
could also be present.  
Each particle is coupled to a Langevin thermostat with
damping constant $\gamma_i$, so that the general equations of
motion read
\begin{equation}\label{langevin}
\begin{aligned}
&\dot {q}_i=  \partial_{p_i} \caH = \frac{ p_i}{ m_i}\equiv  v_i \,,\\
&\dot {p}_i= -\partial_{q_i} \caH + f_i - \gamma_i  p_i +  \xi_i\,.
\end{aligned}
\end{equation}
Here, the $ \xi_i$ represent Gaussian white noise with correlation 
$\mean{ \xi_i(t) \xi_j(t')}= 2 D_{ij} \delta(t-t') $. 
We first consider the case
of independent heat baths in local equilibrium at temperature $T_i$,
for which the fluctuation-dissipation theorem implies a diagonal
diffusivity matrix $D_{ij} = m_i \gamma_i k_{\rm B}T_{i} \delta_{ij}$. In section~\ref{sec:modes} 
we will show an example of a non-diagonal temperature matrix emerging for the normal modes of
coupled oscillators. Note that a space-dependent noise is included in this
formalism, since $T_i$ may be a continuous function $T(q_i)$ of the particle position.

Our results directly derive from the formula for the
time derivative of the average of any state observable $\caO(t)$, 
\begin{align}\label{ave}
  \frac{\diff}{\diff t} \mean{\caO}= \mean{\genL \caO},
\end{align}
where $\genL$ is the backward generator of the dynamics. 
For the Langevin equations~\eqref{langevin} it can be derived with It\^o's
formula~\cite{gardiner} and is given by
\begin{equation}
\genL\!=\!\sum_{i=1}^{N d} \! \left[
\frac{p_i}{m_i} \partial_{q_i} + \left( f_i - \partial_ {q_i} \caH - \gamma_i p_i \right) \partial_{p_i}\! 
+ \sum_{j=1}^{N d}\! D_{ij} \partial_{p_i}\partial_{p_j}\! \right]\!.
\end{equation}
For example, a set of relations emerges from the position-momentum observable $\caO=p_i q_i$
~\footnote{This derivation is formally identical to the one employed in most quantum mechanics textbooks, e.g.~\cite{schiff}. Indeed, $\mean{\genL \caO}=\mean{ \caO \genW}=\mean{ \caO \genW-\genW \caO }=\mean{ [\caO ,\genW]}$. This relation employs, in order, the definition of the generator of forward time evolution $\genW$, the normalization of probabilities, and the definition of the commutator.
 The correspondence between $\genW$ and the quantum generator of time evolution $-\frac{\text{i}}{\hbar} \mathcal{H} $ then gives $\mean{ [\caO ,\genW]}=-\frac{\text{i}}{\hbar}\mean{[\caO, \mathcal{H}]} $. }.
Plugging it into \eqref{ave}, we obtain
\begin{align}\label{pq}
\frac{\diff}{\diff t} \mean{p_i q_i}= \mean{\frac{p_i^2}{m}} 
+\mean{ \left( f_i - \partial_ { q_i} \caH - \gamma_i p_i \right)q_i}.
\end{align}
Using then  
$\mean{p_i q_i}=  m_i \mean{\dot q_i q_i}= \frac 1 2 \frac{\diff}{\diff t}\mean{m_i q_i^2}$
and removing all time derivatives by the assumption of stationarity,
this is turned into a MVE for the conjugated pairs $q_i,p_i$:
\begin{equation}\label{virialin}
\mean{\frac{p_i^2}{m_i}} = \mean{ \left( \partial_{ q_i} \caH - f_i \right)q_i}.
\end{equation}  
The virial theorem follows by applying
$\sum_{i=1}^{N d}$ to both sides of~\eqref{virialin}. 
The equipartition theorem is recovered in equilibrium ($f_i=0$, ${T_i=T}\;\forall i$),
where averages may be performed with the Boltzmann weight $\exp(-\frac{\caH }{k_{\rm B} T})$
and all terms in~\eqref{virialin} are equal to $k_{\rm B}T$.

\begin{figure}[!t]
\center
\begin{tabular}{c}
\includegraphics[angle=0,width=0.3\textwidth]{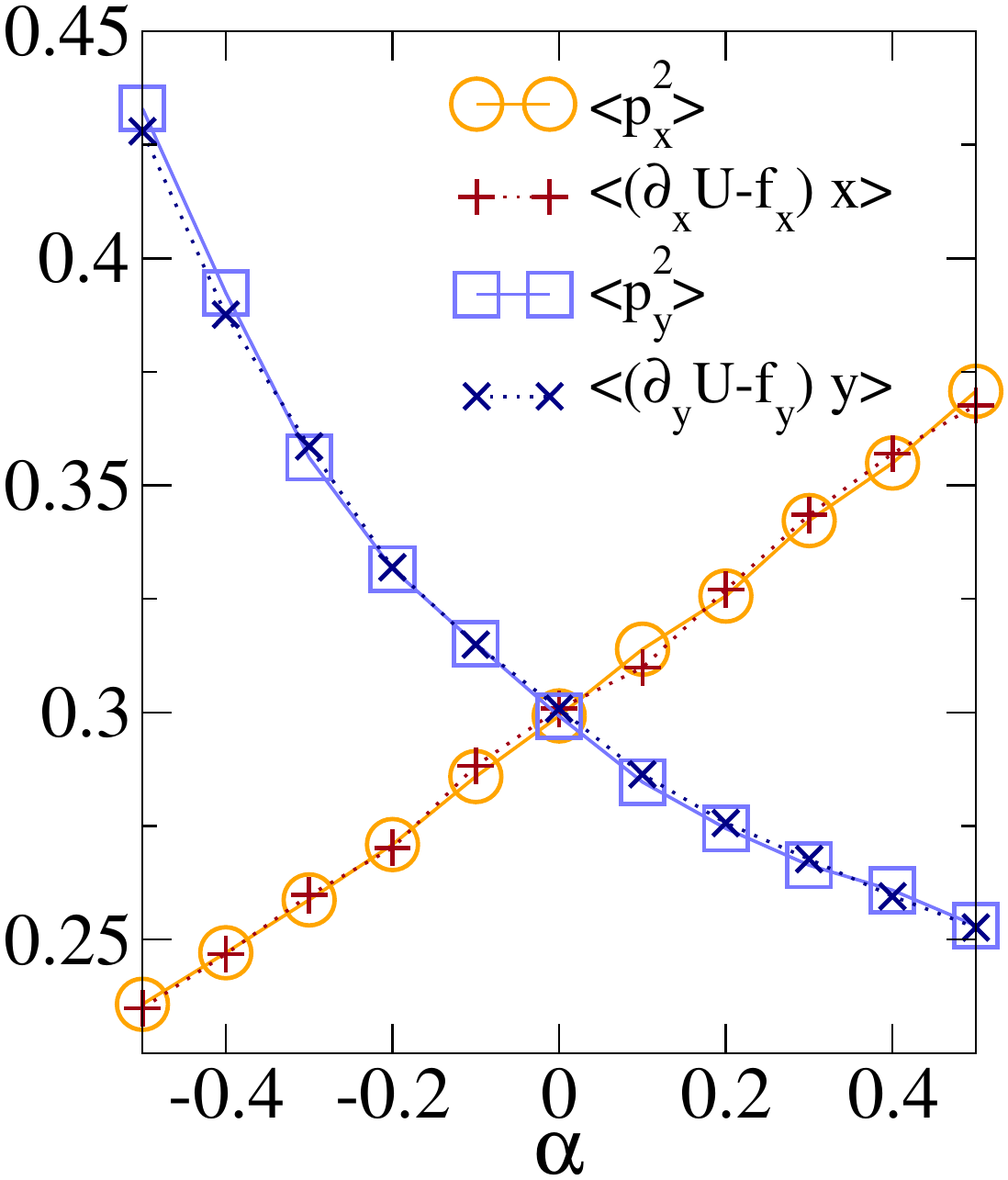}
\includegraphics[angle=0,width=0.38\textwidth]{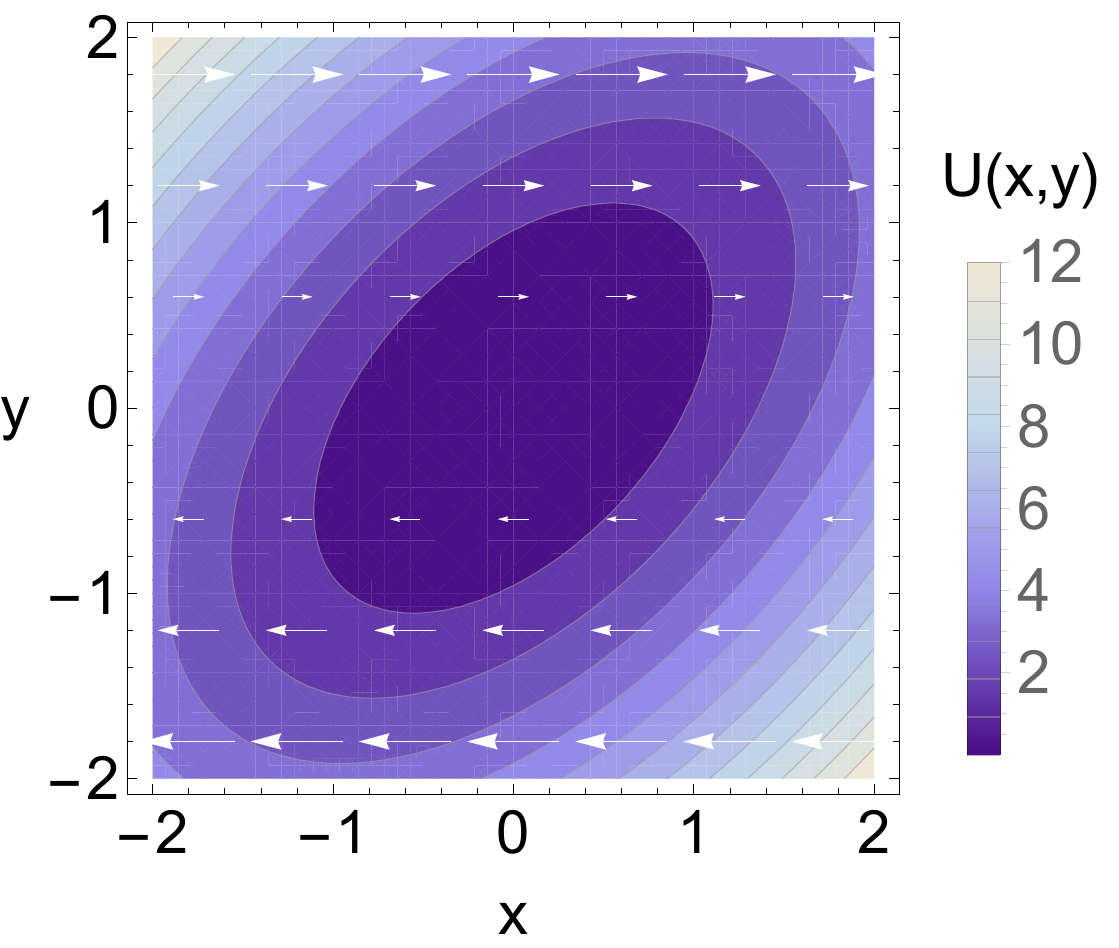}
\end{tabular}
\caption{MVE for a two-dimensional particle, for $T=0.3$ ($k_{\rm B}=1$)
and $\gamma=0.2$, subject to the potential and shear forces sketched on the right
($\alpha>0$ in the example, see the text).}
\label{fig:tilt}
\end{figure}

As a basic exemplification, consider a unit-mass particle moving in the 
$\vec q = (x,y)$ plane, subjected to the potential $U(x,y) = x^2 - x y + y^2$ and
to the nonconservative shear force $\vec f= \alpha\,y\,\vec e_x$ 
($\vec e_x$ being the $x$-axis versor) \footnote{{Throughout the text we employ $\alpha$ as a dimensional constant measuring the departure from equilibrium.}}. 
In Fig.~\ref{fig:tilt}(a) we display a numerical 
validation of the MVE~\eqref{virialin}. 
Note that energy equipartition 
--- with virial and (twice) kinetic contributions 
amounting to $k_{\rm B}T$ --- is achieved  only 
in equilibrium (for $\alpha=0$).
This example also
illustrates that specific degrees of freedom can still be cooled
down under nonequilibrium conditions (for $\alpha\neq 0$)~\cite{fal15.b},
despite energy being constantly supplied to the
particle.
%
%
%
%
%
%
\section{Nonequilibrium equations of state}\label{sec:e-s}
Switching to the observable $\caO= p_i^2$, 
Eq.~\eqref{ave} provides
\begin{align}\label{pp}
\frac{\diff}{\diff t} \mean{p^2_i}= 2\mean{ \left( f_i - \partial_ { q_i} \caH - \gamma_i p_i \right)p_i} +2\mean{D_{ii}}.
\end{align}
Here 
$\langle J_i\rangle = \left\langle( f_i - \partial_ { q_i} \caH ) \frac{p_i}{m_i}\right\rangle$
is recognised as the average heat flow into the $i$-th reservoir,
and in a steady state one gets the Harada-Sasa formula~\cite{har05,lip14}
\begin{align}\label{dissip}
\mean{J_i}  &= \gamma_i\left( \mean{\frac{p_i^2}{m_i}} -  
k_{\rm B}\mean{T_{i}}\right)\,.
\end{align}
Combining now the MVE \eqref{virialin} with \eqref{dissip}, we find
\begin{align}\label{dissip-2}
\frac{\mean{J_i}}{\gamma_i} + k_{\rm B}\mean{T_{i}}  &= \mean{ \left( \partial_{ q_i} \caH - f_i \right)q_i} .
\end{align}
If the system is in thermal equilibrium, then $\mean{J_i} = 0\;\forall i$, 
and~\eqref{dissip-2} constitutes the standard starting point for deriving equations of
state. 
For $N$ interacting particles within a container of
volume $\caV$ it is useful to separate the contribution of the {external conservative
 forces $\vec F_{\rm ext}$ (comprising confining wall forces $\vec F_{\rm w}$, gravity, etc.)} 
from that of the inter-particle
interactions $\vec F_{\rm int}$. The sum over all degrees of freedom 
of $\langle \partial_{ q_i} \caH  \,q_i\rangle$ gives both the {\em external} virial 
{
as a function of the pressure $P$,
\begin{align}\label{C_ext}
-\sum_{i=1}^{N d}\mean{F_{{\rm ext},i}\,q_i} = P \caV d,
\end{align}
 and the {\em internal} virial
$C_{\rm  int} = -\sum_{i=1}^{N d} \mean{F_{\text{int},i}\, q_i}$~\cite{gallavotti}.}
Under equilibrium conditions, from~\eqref{dissip-2} thus descends
\begin{align}\label{e-s-1}
Nk_{\rm B}T  &= P \caV + C_{\rm int}/d,
\end{align}
which can for example be used to derive the van der Waals equation \cite{pathria}. 
Since the above deduction only relies 
on the homogeneity and isotropy of the
system, it is {directly} applicable to
nonequilibrium situations that enjoy these symmetries. 
We may think about
systems with equal particles and homogeneous dissipation
($T_i=T$ and $\;\gamma_i=\gamma\;\forall i$), say. 
As in the case of Fig.~\ref{fig:tilt}(a), nonequilibrium
steady states are maintained by the action of {the nonconservative forces, which contribute in \eqref{dissip-2} the additional nonequilibrium virial term $C_{\rm ne} \equiv-\sum_{i=1}^N\mean{ f_i q_i}$. Different cases should be distinguished, depending on the nature of  $f_i$. If $f_i$ is an external driving, such as the shear force of section \ref{sec:Langevin}, $C_{\rm ne}$ combines with the conservative external forces to produce the pressure,
\begin{align}
-\sum_{i=1}^{N d}\mean{(F_{{\rm ext},i} + f_i)\,q_i}  = P \caV d.
\end{align}
Hence the equation of state \eqref{e-s-1} is generalised to 
\begin{align}\label{e-s-2}
Q^* + N  k_{\rm B}T d &= P \caV d + C_{\rm int},
\end{align}
where $Q^* = \mean{\sum_i J_i }/ \gamma$ is the total heat flowing
into the reservoir during a characteristic time $1/\gamma$.
On the other hand, if $f_i$ is a dissipative interaction force between particles (e.g.~describing binary inelastic collisions in granular gases~\cite{bril}), then \eqref{C_ext} holds true and the nonequilibrium virial $C_{\rm ne}$ figures explicitly in the equation of state
\begin{align}\label{e-s-2}
Q^* + N  k_{\rm B}T d &= P \caV d + C_{\rm int} + C_{\rm ne}.
\end{align}
Notice that these equations of state include not only equilibrium thermodynamic variables
but also an unusual heat-flow contribution $Q^*= \mean{\sum_i f_i \dot q_i}/\gamma$,
which stems solely from the nonconservative driving because stationarity 
implies $\mean{\sum_i  \dot q_i \partial_{q_i} \caH}=\frac{\diff }{\diff t}\mean{U}=0$.}

\begin{figure}[!t]
\center
\begin{tabular}{c}
\includegraphics[angle=0,width=0.45\textwidth]{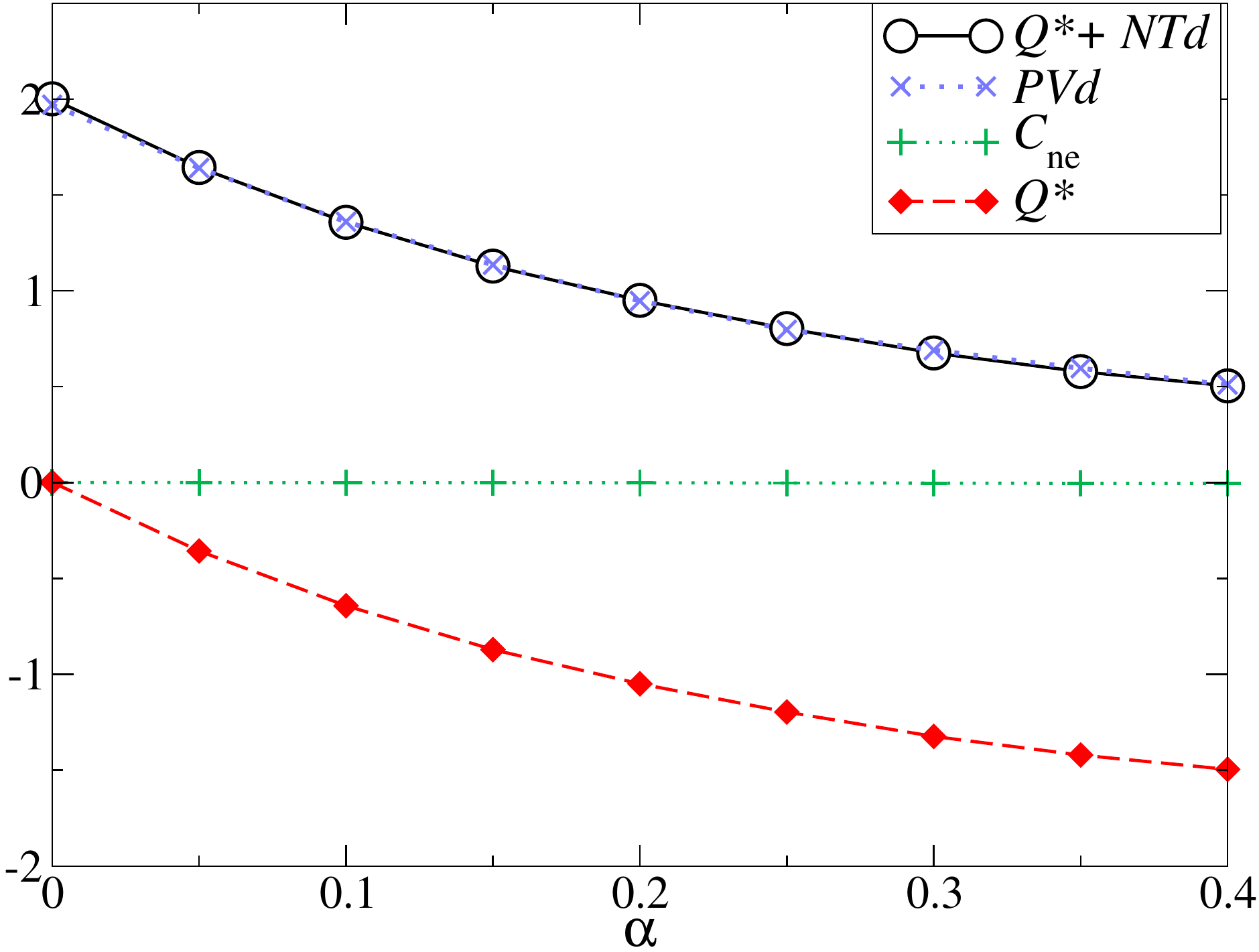}
\end{tabular}
\caption{
  Contributions in~\eqref{e-s-2} for a single particle ($N=1$) at
  temperature $T=1$ (natural dimensionless units). The negative sign of the
  heat flowing into the reservoir, $Q^*<0$, is consistent with a positive
  heat absorbed by the system when solid friction dissipates energy ($\alpha>0$).
}
\label{fig:state-eq}
\end{figure}

As a simple illustration of the role of $Q^*$, 
consider $N$ independent particles with unitary mass, again in the
$xy$-plane.
Each particle is subjected to a Langevin bath of uniform temperature $T$, 
to a confining potential $U_{\rm w}(x,y)=\frac 1 {12}(x^{12}+y^{12})$ 
(thus $\vec F_{\rm w} = -\vec \nabla U_{\rm w}$), 
and to an additional solid friction $\vec f=-\alpha\,\vec v/|\vec v|$
of constant magnitude $\alpha\geq0$~\cite{gnoli13, touchette10}. 
In the presence of this non-conservative friction, a steady state is generated in
which heat is continuously taken from the Langevin bath and delivered
to the substrate. However, the symmetry of the problem implies that 
$C_{\rm ne}$ is zero.
In view of the particles' mutual independency, also $C_{\rm int}$ is exactly
zero, and each of the remaining terms in~\eqref{e-s-2} amounts to $N$ times
the single-particle contribution.
In Fig.~\ref{fig:state-eq} we display each term in~\eqref{e-s-2} 
for single-particle simulations of the system at
various $\alpha$.
As depicted, $Q^*=0$ in equilibrium ($\alpha=0$), while 
$Q^*$ is negative  in nonequilibrium ($\alpha > 0$) and gives important contributions that guarantee the
validity of the equation of state.

\section{Overdamped dynamics}\label{sec:over}

If one considers time scales much larger than the characteristic relaxation times of momenta, 
i.e.~$\gamma_i \diff t  \to \infty $~\cite{gardiner}, then ${\mean{J_i}/\gamma_i \to 0}$ and~\eqref{dissip-2} reduces to the overdamped MVE~\footnote{To avoid the issues related to the interpretation of the overdamped stochastic equations hereafter we consider additive noise only.}
\begin{equation}
\label{virialin-o}
k_{\rm B}T_i=\mean{\left(\frac{\partial U}{\partial  q_i} - f_i \right) q_i}\,.
\end{equation} 
This corresponds to \eqref{virialin} after the substitution 
$\mean{p_i^2/m_i} \mapsto k_{\rm B}T_{i}$, as it should be expected, 
since momentum is instantaneously thermalised by its own thermal bath in the overdamped limit.
Of course, this relation can be derived directly by taking the overdamped limit of the
diffusion equations~\eqref{langevin}:
\begin{equation}\label{overdamped}
 \dot {q}_i= \mu_i\left(-\partial_{q_i} U + f_i\right)+ \hat \xi_i\,,
\end{equation}
where $\mu_i=(m_i \gamma_i)^{-1}$ is the mobility,
$\mean{\hat \xi_i(t) \hat \xi_j(t')} = 2 \hat{D}_{ij}\,\delta(t-t')$
with
$ \hat{D}_{ij}= \mu_i k_{\rm B}T_{i} \delta_{ij}$, 
and the Hamiltonian $\caH$ boils down to the potential energy $U$. 
The backward generator of the dynamics becomes
${\genL}=\sum_i \mu_i(f_i -\partial_{q_i} U) \partial_{q_i}+ \sum_{ij} \hat{D}_{ij}\partial_{q_i}\partial_{q_j}$,
and $\caO = q_i^2$ is the appropriate observable to plug in~\eqref{ave} to retrieve \eqref{virialin-o}.

{
These results hold under the assumption that the dissipative force $f_i$ acts effectively on time scales much longer than $1/\gamma_i$. If instead $f_i$ is of order $O(\gamma_i) $, energy dissipation interferes with the thermalization process of momenta, so that  $\mean{p_i^2/m_i} \neq k_{\rm B}T_{i}$. For example, a solid friction (see section \ref{sec:e-s}) of order $O(\alpha_i) \sim O(\gamma_i \mean{|\vec p_i|})$ renders    \eqref{dissip} in the form
\begin{align}
\mean{\frac{p_i^2}{2 m_i}}= k_{\rm B} T_i  - \frac{\alpha_i\mean{|\vec p_i|}}{\gamma_i m_i},
\end{align}
and thus yields an overdamped MVE which features nonequilibrium corrections to the bath temperature, of the form
\begin{equation}
k_{\rm B} T_i- \frac{\alpha_i\mean{|\vec p_i|}}{\gamma_i m_i}=\mean{\left(\frac{\partial U}{\partial  q_i} - f_i \right) q_i}\,.
\end{equation}}

{ 
Active Brownian particles  (see more details in the next section) can be taken as another example. In the overdamped limit, they are often modeled as colloidal particles driven by a propulsion force $f_{\text{p},i}$ that is counterbalanced by an associated viscous drag force $-\alpha_i p_i$. Together they combine into the non-equilibirium force $f_i= -\alpha_i p_i + f_{\text{p},i}$. 
If the friction forces are comparable in magnitude, that is 
$\alpha_i/\gamma_i=\,$const in the limit $\gamma_i \to \infty$, 
equation  \eqref{dissip}  in the overdamped limit  reads
\begin{align}
\mean{\frac{p_i^2}{2 m_i}}= \frac{k_{\rm B} T_i}{1+ \frac{\alpha_i}{\gamma_i}},
\end{align}
which implies a renormalised temperature for the overdamped MVE
\begin{align}
 \frac{k_{\rm B} T_i}{1+ \frac{\alpha_i}{\gamma_i}}=\mean{\left(\frac{\partial U}{\partial  q_i} - f_i \right) q_i}.
 \end{align}
  }

\section{Overdamped active matter}
Active Brownian particles are often employed as an {overdamped} model for the collective behaviour 
of motile bacteria and self-propelled colloids~\cite{mar13}.
Their phase behaviour is currently much studied~\cite{tak14,ginot15,sol15, sol15b,yan15}.  
In this regard, the utility of the virial theorem was pointed out in Ref.~\cite{win15}.
Here we {fully exploit the generalised virial theorem and} show how our approach leads to a pressure equation for active particles confined 
by hard walls of arbitrary geometry.

We describe an ensemble of identical active Brownian spheres moving in a
two-dimensional volume $\caV$ in terms of their positions 
$\vec r_i=(x_i,y_i)$ and velocity orientations $\theta_i$
(hence, $\{\vec q \} = \{\vec r, \theta\}$). 
Their overdamped equations of motion are
\begin{align}
\dot{\vec r}_i &=  v_0 \vec u(\theta_i)+ \mu\vec F_{\rm w}(\vec r_i) 
                  + \sum_{j \neq i} \mu\vec F_{\rm int}(\vec r_i- \vec r_j) 
                  +  \hat{\vec \xi}^{(r)}_i ,\nonumber\\ 
\dot{\theta}_i &= \hat\xi^{(\theta)}_i .
\end{align}
The active velocity of modulus $v_0$ is directed along the unit vector 
$\vec u(\theta_i)=(\cos \theta_i, \sin \theta_i)$, 
and can be {formally} interpreted as another realization of the nonconservative force $\vec f$
that breaks detailed balance. {Specifically, $\vec f= -\alpha \vec p + \vec{f}_{\text{p}}$ (see section \ref{sec:over}) so that $v_0 \vec u(\theta_i) = \mu \vec f_{\text{p}}$ with $\mu= [m(\gamma+ \alpha)]^{-1}$. }
Each particle experiences the others through the two-body force $\vec F_{\rm int}$. 
No special symmetry is assumed for the confining hard walls acting via $\vec F_{\text{w}}(\vec r_i)$ 
at the container surface $\caS$. The Gaussian translational noise $\hat{\vec \xi}^{(r)}_i$ 
is characterised by
$\mean{\hat{\vec \xi}^{(r)}_i(t) \hat{\vec \xi}^{(r)}_j(t')} = 2 \mu k_{\rm B}T \delta_{ij} \vec 1 \,\delta(t-t')$
and the Gaussian rotational noise $\hat\xi^{(\theta)}_i$ by
$\mean{\hat\xi^{(\theta)}_i(t) \hat\xi^{(\theta)}_j(t')} = 2 \hat{D}^{(\theta)}\delta_{ij}\,\delta(t-t')$.
The backward generator $\genL$ is thus
\begin{align}\label{dO}
\genL= \sum_{i=1}^{N} \bigg[
\Big( 
v_0 \vec u(\theta_i)+ \mu\vec F_{\rm w}(\vec r_i) + \mu\sum_{j \neq i} \vec F_{\rm int}(\vec r_i- \vec r_j)  
\Big) 
 \cdot
\nabla_{\vec r_i}
 +  \mu k_{\rm B}T \nabla^2_{\vec r_i} +\hat{D}^{(\theta)}\partial^2_{\theta_i} \bigg],
 \end{align}
and the choice of the observable $\caO=\vec r_i^2$ in \eqref{ave} yields the
overdamped MVE
\begin{align}\label{virialx}
2k_{\rm B}T=
-&\bigg<\Big( \frac{1}{\mu}v_0 \vec u( \theta_i)+ \vec F_{\rm w}(\vec r_i)+ \sum_{j \neq i} \vec F_{\rm int}(\vec r_i- \vec r_j) \Big) \cdot \vec r_i \bigg>.
\end{align}
Summing over all particles, the term containing $\vec F_{\rm w}$ becomes
the external virial which is most conveniently expressed in terms of the local particle density, $\rho(\vec r) =\mean{ \sum_i^N \delta(\vec r - \vec r_i)}$, as
\begin{align}\label{rho-1}
-\sum_{i=1}^N \mean{\vec F_{\text{w}}(\vec r_i) \cdot  \vec r_i} = -\int_{\caV} \diff \vec r \vec F_{\text{w}}(\vec r) \cdot \vec r \rho(\vec r).
\end{align}
Since the local stress tensor $\vec P$ is defined by the steady-state equation expressing momentum conservation ~\cite{irving50},
\begin{align}\label{hydro}
\nabla_{\vec r} \cdot \vec P(\vec r)=\vec F_{\text{w}}(\vec r) \rho(\vec r),
\end{align} 
an integration by parts of \eqref{rho-1} yields $
-\sum_{i=1}^N \mean{\vec F_{\text{w}}(\vec r_i) \cdot  \vec r_i} = 2 \bar P_\caV \caV$.
Here $\bar P_\caV$ is the volume-averaged pressure, defined through the trace of the pressure tensor $\bar P_\caV \equiv \frac {1}{2 \caV} \int_{\caV} \diff \vec r \textrm{Tr} \vec P (\vec r)$. One expects the pressure to be non-uniform due to particle aggregation at the boundaries \cite{yang14, sma15} and phase separation~\cite{bia15} in the presence of activity, unless highly symmetric geometries are considered~\cite{sol15}. 
{Note that in the momentum balance \eqref{hydro} the only external force is the wall interaction. Consistently with the assumption of a constant active speed $v_0$, the self-propulsion force and the corresponding fluid friction are taken to balance each other, hence are not included in the righthand side of  \eqref{hydro}.}

For the special case of hard walls, the external virial is proportional to the surface-averaged density 
$\bar \rho_\caS$,   
namely $-\sum_{i=1}^N \mean{\vec F_{\text{w}}(\vec r_i) \cdot  \vec r_i} = 2 \caV k_{\text{B}} T  \bar \rho_\caS$~\cite{hend86, pow85}. Moreover, inter-particle interactions do not contribute to the momentum flux at the wall, so that the surface-averaged pressure $\bar P_\caS$ can only have a kinetic contribution~\cite{pow85, fish64}, $ \bar P_\caS= k_{\text{B}}T \bar \rho_\caS$. {The latter is an equilibrium result that can be employed here, since in this overdamped description momenta are assumed to be thermalised at the temperature $T$. This follows from the choice of the translational noise's correlation. }  
Therefore one arrives at the important result that the external virial gives the mean force per unit area exerted on the container,
\begin{align}\label{rho}
-\sum_{i=1}^N \mean{\vec F_{\text{w}}(\vec r_i) \cdot  \vec r_i} = 2 \bar P_\caS \caV.
\end{align}

In the bulk, the interaction term in~\eqref{virialx} gives a contribution
analogous to the
corrections to the ideal gas pressure in an equilibrium
system. 
Indeed, for large $N$,
\begin{align}
& \sum_{i,j \neq i}\mean{ \vec F_{\rm int}(\vec r_i- \vec r_j) \cdot \vec r_i} =  -\frac{N^2}{2 \caV^2}  \int_{\caV} \diff \vec r' \int_{\caV} \diff \vec r'' r \frac{\partial U_{\rm int}}{\partial r} g(\vec r',\vec r'') ,
\end{align}
where $\vec F_{\rm int}=-\nabla U_{\rm int}$, $r\equiv |\vec r'- \vec r''|$, and $g$ is the
nonequilibrium pair density correlation function. 
In general, $g$ cannot be reduced to a function of the relative pair position,
since the system is inhomogeneous~\cite{pli86}.  
The explicit nonequilibrium contribution in~\eqref{virialx} 
(the term containing $v_0$) gives rise to the so-called swim 
pressure~\cite{yang14,bia15}. Using~\eqref{ave}, 
this time with $\caO= \vec r_i \cdot \vec u(\theta_i)$, 
and summing, we readily  obtain 
\begin{align}
v_0\hat{D}^{(\theta)} \sum_i \mean{\vec r_i \cdot \vec u(\theta_i) }
= 
v_0^2+ v_0 \mu \sum_i\mean{ \vec F_{\rm w}(\vec r_i) \cdot \vec u(\theta_i)}
 + v_0 \mu \sum_{i,j \neq i}  \mean{\vec F_{\rm int}(\vec r_i- \vec r_j) \cdot \vec u(\theta_i)}. 
\end{align}
The first average on the r.h.s. involves the particle polarization at
the wall, while the second one represents 
the correlation between interactions and polarization. 
The constant term $v_0^2$ is an enhancement of the kinetic ``ideal gas''
contribution due to the particles' activity.
Putting everything together, we obtain the equation of state 
\begin{align}\label{eqABP}
 \bar P_\caS \caV =&N k_{\rm B}T -\frac{N^2}{4 \caV^2}  \int_{\caV} \diff \vec r'
 \int_{\caV} \diff \vec r'' r \frac{\partial U_{\rm int}}{\partial r} g(\vec r, \vec r')
 \\
& + \frac{v_0^2}{2\mu\hat{D}^{(\theta)}}
+ \frac {v_0}{2\hat{D}^{(\theta)}}  \sum_{i,j \neq i}\mean{ \vec F_{\rm int}(\vec r_i- \vec r_j) \cdot \vec u(\theta_i)}
+ \frac {v_0}{2\hat{D}^{(\theta)}} \sum_i \mean{\vec F_{\rm w}(\vec r_i) \cdot \vec u(\theta_i)} \nonumber.
\end{align}
This result is valid irrespective of the confining geometry, thus extending the results of ~\cite{win15} and substantiating the numerical evidence for the equality of (average) wall and bulk pressure \cite{yang14, sma15}. 
{It is worth mentioning that it is currently under debate  \cite{yan15,sol15b} whether \eqref{eqABP} deserves the name of state equation, given that the active particle pressure appears to depend explicitly on the interactions with the boundaries and not only on bulk properties (temperature, density, etc.) as it would happen in equilibrium.}

\section{Normal modes of coupled oscillators}\label{sec:modes}
The derivation of the MVE does not rely on the diagonality of the the matrix
$D_{ij}$, that is \eqref{virialin} also holds for systems in which 
the noise components are cross-correlated. 
An instance of such a situation is offered by the
analysis of the normal modes
of a system with local reservoirs.
For harmonic lattices~\cite{fal15}, 
depending on the details of the forcing and on boundary conditions, 
the energy stored in long wavelength vibrational modes
may be either enhanced or reduced compared to the average.
Here, we illustrate the MVE in modes' space
for a one-dimensional chain of $N$ point masses coupled with quadratic-quartic interactions, 
thus going beyond the harmonic approximation. The stochastic equation of the normal modes,
{obtained by applying a linear transformation to the equation \eqref{langevin} for the oscillators' position and velocity \cite{inp}, is}
\begin{equation}\label{LangevinX2}
\ddot X_k= -\gamma \dot X_k - \omega_k^2 X_k - \epsilon \sum_{l,r,s}\caB_{klrs}X_l X_r X_s+ \eta_k,
\end{equation}
where $\omega_k^2$ is
the squared eigenfrequency of the $k$-th mode
and  $\epsilon \caB_{klrs}$
is a  tensor that emerges from the quartic interactions.
The noise terms $\eta_k$,
\begin{align}
\mean{\eta_k(t) \eta_l(t')} & = 2 \gamma k_{\rm B}\caT_{kl}  \delta(t-t'),
\end{align}
are mutually correlated, as
quantified by a symmetric matrix $\caT_{kl}$ of {\em mode temperatures} ~\cite{fal15.b},
which is in general not diagonal unless the system is in equilibrium.

\begin{figure}[!t]
\center
\begin{tabular}{ll}
(a)&(b)\\
\includegraphics[width=0.5\textwidth]{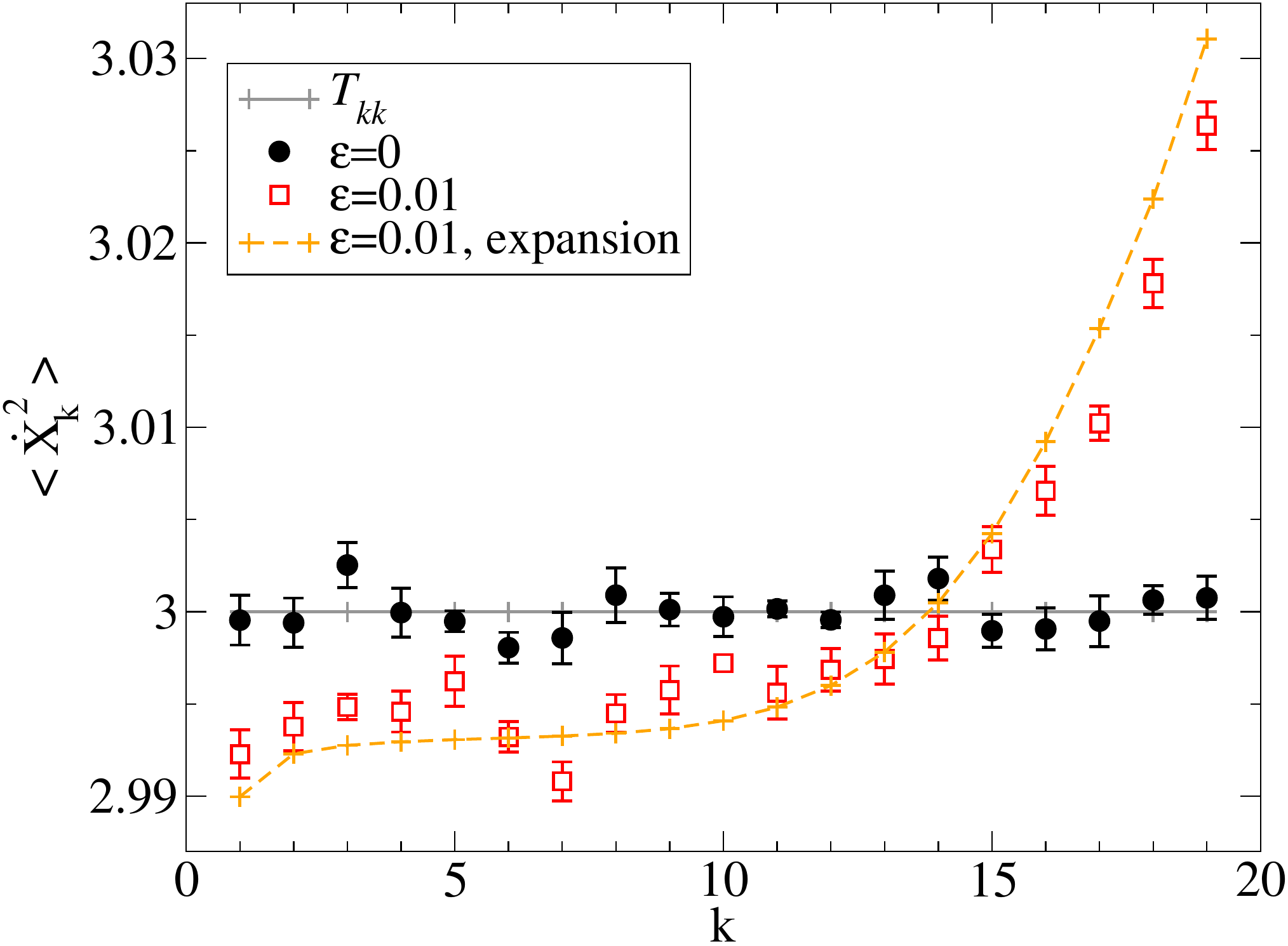}
&
\includegraphics[width=0.5\textwidth]{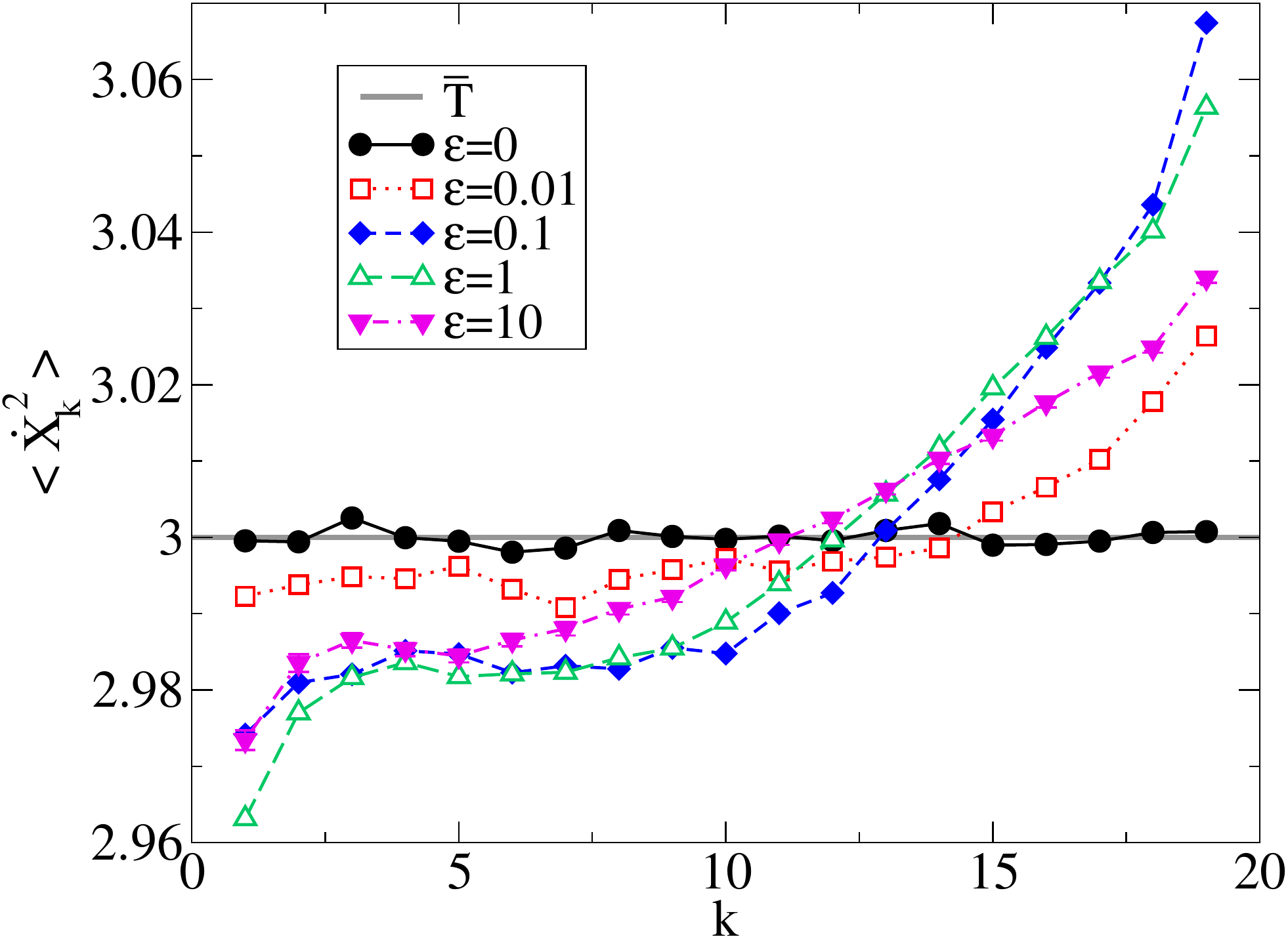}
\end{tabular}
\caption{
Kinetic energy of the normal modes for a chain of
$N=20$ unit masses coupled via quadratic-quartic potential (harmonic constant $\kappa=1$, 
quartic $\epsilon = 0, \dots, 10$) and immersed in heat baths characterised by the (global) 
friction constant $\gamma=0.1$ and the local temperatures $T_i$, which grow
linearly with $i$ from $T_0=1$ to $T_{N-1}=5$ (in natural dimensionless units). 
(a) Comparison between the analytic expansion \eqref{perturb} (\textcolor{orange}{+}) and the numerically estimated $\mean{\dot X_k^2}$ ({$\square$}) obtained by integration of the oscillators' stochastic dynamics. 
(b) Numerically estimated mode kinetic energies also for strongly anharmonic chains. Errors are of the order of symbol sizes.
}
\label{fig:dotX}
\end{figure}

Without the anharmonic coupling,
$\epsilon=0$, the average kinetic and potential energy of the
modes satisfy 
\begin{align}\label{MSmode}
\mean{\dot X_k^2}=\omega^2_k\mean{X_k^2}=k_{\rm B}\caT_{kk}\,,
\end{align}
where the first equality is analogous to~\eqref{virialin},  
and the second amounts to~\eqref{dissip-2} specialised to the
present analysis. 
Notice that the kinetic and potential energy coincide for a given mode,
but they  in general differ for different modes, thus breaking
full equipartition.  
With $\epsilon\neq0$ the modes' dynamics is coupled via the tensor
$\caB_{klrs}$ and the MVE~\eqref{virialin} becomes
\begin{equation}
\label{virialinX}
\mean{\dot X^2_k} = \omega_k^2 \mean{X_k^2} + 
\epsilon \sum_{l,r,s} \caB_{klrs} \mean{X_k X_l X_r X_s}
\end{equation}
containing no explicit sign of the non-diagonal $\caT_{kl}$, as anticipated above. 
{Similarly, the heat-flux equation~\eqref{dissip} becomes
\begin{align}\label{virialinXdot}
\mean{\dot X_k^2} &= k_{\rm B} \caT_{kk} + \frac{\epsilon}{\gamma} \sum_{l,r,s}  \caB_{klrs} \mean{\dot X_k X_l X_r X_s}.
\end{align}
This represents the perfect starting point to study 
perturbative corrections to  mode energies, given the Gaussian statistics of the $X_k$'s for $\epsilon=0$. 
The last term disappears in equilibrium ($T_i=T\, \forall i $) where the modes' position and velocity are on average uncorrelated, so that the equipartition for velocities $\mean{\dot X_k^2} = k_{\rm B} T$ results from \eqref{virialinXdot}. 
Differently, the non-zero heat flux present under non-equilibrium conditions modifies the mode kinetic energy in \eqref{virialinXdot}. For small $\epsilon$ we can expand \eqref{virialinXdot} as
\begin{align}\label{perturb}
\mean{\dot X_k^2} &= k_{\rm B} \caT_{kk} + \frac{3 \epsilon}{\gamma} \sum_{l,r,s}  \caB_{klrs} \langle\dot X_k X_l \rangle_{\epsilon=0}\mean{X_r X_s}_{\epsilon=0} +O(\epsilon^2).
\end{align}
Here we used the symmetry of the tensor $\caB$ together with Wick's theorem to break up the Gaussian correlations $\mean{\dots}_{\epsilon=0}$ evaluated in the harmonic system~\cite{fal15.b}. An illustration of \eqref{perturb} is provided in Fig.~\ref{fig:dotX}(a) for a one-dimensional lattice with fixed boundaries immersed in a linear temperature profile. For $\epsilon=0$, due to the symmetry in the $T_i$'s and in the boundary conditions, the modes enjoy a peculiar full energy equipartition \cite{fal15.b} at the average temperature 
$ \caT_{kk} = \overline T \equiv  \frac 1 N \sum_{i=0}^{N-1} T_i $.  The anharmonic corrections allow energy to leak into the higher, more localised modes. The same qualitative behaviour is found numerically for increasing values of $\epsilon$ (Fig.~\ref{fig:dotX}b).
This suggests that the energy repartition among modes is robust against the introduction of non-linearities and fairly well approximated by a first order perturbative calculation.
Note that the total energy is insensitive to $\epsilon$, namely 
$\sum_{k=0}^{N-1} \mean{\dot X_k^2} = k_{\rm B} \sum_{k=0}^{N-1} \caT_{kk}$ $\forall \epsilon$, 
since the total heat flux appearing in \eqref{virialinXdot} is identically zero under stationary conditions, thanks to the potential nature of the interaction:
\begin{align}
 \sum_{k,l,r,s}  \caB_{klrs} \mean{\dot X_k X_l X_r X_s} = \sum_{k=0}^{N-1} \mean{\dot X_k (\partial_k U - \omega_k^2 X_k)}= \frac{d}{dt} \left(\mean{U}- \sum_{k=0}^{N-1}\frac{\omega_k^2 }{2} \mean{X_k^2}\right)=0.
\end{align}
}

\section{Deterministic thermostats}
The relations derived above for stochastic inertial systems remain valid in
the zero-noise limit, where the dynamics becomes
deterministic. Stationarity is then ensured by coupling the system to suitably defined thermostats. 
Examples are Nos\'e-Hoover thermostats, where extra degrees of freedom act as frictional couplings 
for the physical ones~\cite{hoover85}.
Similarly to Langevin dynamics, they guarantee canonical thermalization in cases of uniform
temperatures, and they lead to non-zero heat
fluxes if different temperatures are imposed on different degrees of freedom of the system. 
For lattices of oscillators interacting only via conservative forces and coupled to Nos\'e-Hoover thermostats at various temperatures, the existence of local energy equipartition is a common assumption needed for the local definition of temperature \cite{lep98}.
So far, it has only been observed in simulations 
for the masses not directly driven by Nos\'e-Hoover thermostats~\cite{gib11}. Here we provide a formal proof.
We consider statistical averages with respect to the invariant density, which, in general, may or may not coincide with time averages. 
Equality is in fact assured by the use of Nos\'e-Hoover chains of thermostats \cite{marty92}.

 The Nos\'e-Hoover dynamics for unit masses is given by
\begin{align}\label{NH}
\dot p_i = -\partial_{q_i}U - \Theta_i \zeta_i p_i, \qquad \dot \zeta_i = 
\frac{1}{\tau^2} \left(\frac{p_i^2}{k_{\rm B}T_{i}} -1\right),
\end{align}
where $\Theta_i$ is an indicator function, which is 1 or 0 depending on whether
the mass $i$ is coupled or not to a thermostat. 
The auxiliary feedback variable $\zeta_i$ aims at thermalizing $p_i$
at the temperature $T_{i}$ on a timescale $\tau$.
The backward generator associated to \eqref{NH} is
\begin{align*}
\genL = &\sum_{i=1}^{N d} \bigg\{ p_i\partial_{q_i} - \partial_{q_i}U\partial_{p_i} +\Theta_i \left[-\zeta_i p_i\partial_{p_i}+\frac{1}{\tau^2} \left(\frac{p_i^2}{k_{\rm B}T_{i}} -1\right) \partial_{\zeta_i}\right]
\bigg\}.
\end{align*}
Following the scheme outlined above, we find the generalised MVE
\begin{equation}
\mean{p_i^2} = \mean{  q_i\, \partial_{ q_i} U} + \Theta_i\mean{\zeta_i p_i q_i},
\end{equation}
which includes the formal justification for the mentioned numerical observation 
of local energy equipartition if restricted to masses without a local thermostat~\cite{gib11},
corresponding to $\Theta_i=0$.
The term 
\begin{align}\label{virialin-NH}
\mean{\zeta_i p_i q_i} = - \frac 1 {2\tau^2} \mean{ \left( \frac{p_i^2}{k_{\rm B}T_{i}} -1 \right) q_i^2 },
\end{align}
stemming from the thermostat's force {(that can be seen as another realization of the non-conservative force $f_i$)}, is identically zero only in equilibrium, 
where momentum and position are uncorrelated and $\mean{p_i^2}=k_{\rm B}T_i$
holds also for the degrees of freedom coupled to thermostats.

\section{Conclusions}
For a wide class of nonequilibrium systems in steady states,
including stochastic and deterministic thermostated dynamics,
we have shown that the kinetic energy
of a given degree of freedom is on average equal to the corresponding
virial of the forces. 
An integration over all degrees of freedom of such MVE
yields the standard (macroscopic) virial theorem and 
a variety of useful results for general nonequilibrium systems.
It is indeed possible to follow the path valid for
equilibrium systems, using the virial theorem as a tool for the
derivation of equations of state that involve pressure, temperature
and other observables.
For inertial systems with dissipative
dynamics, this leads to an intriguing relation between the virial, the temperature of the heat baths, and the heat flux into them.
Similarly, for active Brownian particles the virial theorem
represents a powerful tool for deducing an equation of state
valid for arbitrary container geometries.
A direct experimental verification of the fundamental mesoscopic
virial relations (underlying all these results) would therefore
be desirable. In boundary driven systems with conservative internal
forces, such verification amounts to checking energy equipartition
between momentum-position type conjugate variables.

\section*{Acknowledgments}
{G.~F. thanks S. Steffenoni for stimulating discussions. G.~F. and K.~K. acknowledge funding by the Deutsche Forschungsgemeinschaft (DFG).}

\section*{References}
\bibliography{virial_NJP}

\providecommand{\newblock}{}
\begin{thebibliography}{10}
\expandafter\ifx\csname url\endcsname\relax
  \def\url#1{{\tt #1}}\fi
\expandafter\ifx\csname urlprefix\endcsname\relax\def\urlprefix{URL }\fi
\providecommand{\eprint}[2][]{\url{#2}}

\bibitem{gallavotti}
Gallavotti G 1999 {\em Statistical Mechanics: A Short Treatise\/} Texts and
  monographs in physics (Springer) ISBN 9783540648833

\bibitem{mar07}
Marc G and {Mc Millan} W~G 2007 {\em The virial theorem\/} (John Wiley \& Sons,
  Inc.) chap~4, pp 209--361

\bibitem{con13}
Conti L, {De Gregorio} P, Karapetyan G, Lazzaro C, Pegoraro M, Bonaldi M and
  Rondoni L 2013 {\em J. Stat. Mech.\/}  P12003

\bibitem{agu15}
{Aguilar Sandoval} F, Geitner M, Bertin E and Bellon L 2015 {\em J. Appl.
  Phys.\/} {\bf 117} 234503

\bibitem{jol11}
Joly L, Merabia S and Barrat J~L 2011 {\em EPL\/} {\bf 94} 50007

\bibitem{fal14}
Falasco G, Gnann M~V, Rings D and Kroy K 2014 {\em Phys. Rev. E\/} {\bf 90}(3)
  032131

\bibitem{fal15}
Falasco G, Saggiorato G and Vulpiani A 2015 {\em Physica A\/} {\bf 418} 94--104

\bibitem{cha64}
Chandrasekhar S 1964 {\em Lectures in Theoretical Physics\/} vol~6 ed Brittin
  W~E (Boulder: University of Colorado Press) p~1

\bibitem{sol15}
Solon A~P, Stenhammar J, Wittkowski R, Kardar M, Kafri Y, Cates M~E and
  Tailleur J 2015 {\em Phys. Rev. Lett.\/} {\bf 114} 198301

\bibitem{gardiner}
Gardiner C~W 2004 {\em Handbook of stochastic methods for physics, chemistry
  and the natural sciences\/} 3rd ed ({\em Springer Series in Synergetics\/}
  vol~13) (Berlin: Springer-Verlag) ISBN 3-540-20882-8

\bibitem{schiff}
Schiff L~I 1968 {\em Quantum Mechanics\/} 3rd ed (New York: McGraw-Hill)

\bibitem{fal15.b}
Falasco G, Baiesi M, Molinaro L, Conti L and Baldovin F 2015 {\em Phys. Rev.
  E\/} {\bf 92}(2) 022129

\bibitem{har05}
Harada T and Sasa S 2005 {\em Phys. Rev. Lett.\/} {\bf 95} 130602

\bibitem{lip14}
Lippiello E, Baiesi M and Sarracino A 2014 {\em Phys. Rev. Lett.\/} {\bf 112}
  140602

\bibitem{pathria}
Pathria R~K 1996 {\em Statistical Mechanics\/} 2nd ed (Oxford: Elsevier)

\bibitem{bril}
Brilliantov N~V and P{\"o}schel T 2010 {\em Kinetic theory of granular gases\/}
  (Oxford University Press)

\bibitem{gnoli13}
Gnoli A, Puglisi A and Touchette H 2013 {\em EPL\/} {\bf 102} 14002

\bibitem{touchette10}
Touchette H, Van~der Straeten E and Just W 2010 {\em J. Phys. A\/} {\bf 43}
  445002

\bibitem{mar13}
Marchetti M~C, Joanny J~F, Ramaswamy S, Liverpool T~B, Prost J, Rao M and
  {Aditi Simha} R 2013 {\em Rev. Mod. Phys.\/} {\bf 85} 1143

\bibitem{tak14}
Takatori S~C, Yan W and Brady J~F 2014 {\em Phys. Rev. Lett.\/} {\bf 113}
  028103

\bibitem{ginot15}
Ginot F, Theurkauff I, Levis D, Ybert C, Bocquet L, Berthier L and
  Cottin-Bizonne C 2015 {\em Physical Review X\/} {\bf 5} 011004

\bibitem{sol15b}
Solon A~P, Fily Y, Baskaran A, Cates M, Kafri Y, Kardar M and JTailleur 2015
  {\em Nat. Phys.\/} {\bf 10} 673--678

\bibitem{yan15}
Yan W and Brady J~F 2015 {\em Soft Matter\/} {\bf 11} 6235

\bibitem{win15}
Winkler R, Wysocki A and Gompper G 2015 {\em Soft Matter\/} {\bf 11} 6680--6691

\bibitem{irving50}
Irving J~H and Kirkwood J~G 1950 {\em J. Chem. Phys.\/} {\bf 18} 817--829

\bibitem{yang14}
Yang X, Manning M~L and Marchetti M 2014 {\em Soft Matter\/} {\bf 10}
  6477--6484

\bibitem{sma15}
Smallenburg F and L\"owen H 2015 {\em Phys. Rev. E\/} {\bf 92}(3) 032304

\bibitem{bia15}
Bialk{\'e} J, Siebert J~T, L{\"o}wen H and Speck T 2015 {\em Phys. Rev.
  Lett.\/} {\bf 115} 098301

\bibitem{hend86}
Henderson J~R 1986 {\em J. Chem. Phys.\/} {\bf 84} 3385--3386

\bibitem{pow85}
Powles J~G, Rickayzen G and Williams M~L 1985 {\em J. Chem. Phys.\/} {\bf 83}
  293--296

\bibitem{fish64}
Fisher I~Z 1964 {\em Statistical theory of liquids\/} vol~28 (University of
  Chicago Press)

\bibitem{pli86}
Plischke M and Henderson D 1986 {\em J. Chem. Phys.\/} {\bf 84} 2846--2852

\bibitem{inp}
Falasco G, Baiesi M and Baldovin F {in preparation\/}

\bibitem{hoover85}
Hoover W~G 1985 {\em Phys. Rev. A\/} {\bf 31} 1695

\bibitem{lep98}
Lepri S, Livi R and Politi A 1998 {\em Physica D\/} {\bf 119} 140--147

\bibitem{gib11}
Giberti C and Rondoni L 2011 {\em Phys. Rev. E\/} {\bf 83} 041115

\bibitem{marty92}
Martyna G~J, Klein M~L and Tuckerman M 1992 {\em J. Chem. Phys.\/} {\bf 97}
  2635--2643

\end{thebibliography}

\end{document}